\documentclass[12pt]{article}
\usepackage{graphics}

\def\eq#1{(\ref{#1})}

\begin{document}

\title{A Simple Regularization of the Smooth Quantum Hydrodynamic Model}

\author{Carl L. Gardner\\
  School of Mathematical \& Statistical Sciences\\
  Arizona State University, Tempe, AZ 85287\\
  carl.gardner@asu.edu\\
  ORCID\# 0000-0001-8967-3356}

\maketitle
\thispagestyle{empty}

\begin{abstract}
  A simple regularization of the smooth quantum hydrodynamic model
  equations to prevent an unstable growing mode is proposed.  The
  regularization involves replacing the spatial derivative of the
  electron density on the right-hand side of the momentum conservation
  equation by using the classical Boltzmann distribution for electron
  density.

  Time-dependent simulations of the resonant tunneling diode to steady
  state using this regularization are presented, which show realistic
  negative differential resistance (the experimental signal of quantum
  resonance) and hysteresis in the current-voltage curve.  The
  simulations are in good agreement with fully quantum mechanical
  simulations of the resonant tunneling diode.

  This note is an addendum to the author's ``Time-dependent numerical
  methods for a regularized quantum hydrodynamic model''
  \cite{regular-QHD}.
\end{abstract}

\section{Introduction}

In this investigation, we apply a simple regularization (to preclude
an unstable growing mode) of the time-dependent smooth quantum
hydrodynamic model \cite{SmoothQHD} to simulating negative
differential resistance\footnote{A region in the current-voltage curve
  where the current {\em decreases}\/ as the voltage {\em increases}.}
(NDR)---the experimental signal of quantum resonance---and hysteresis
in the current-voltage curve of the resonant tunneling diode.

We solve the time-dependent smooth quantum hydrodynamic (QHD)
equations by using a mixture of hyperbolic, parabolic, and elliptic
partial differential equation methods \cite{regular-QHD}: (i) the
underlying hyperbolic gas dynamical part of the transport equations is
solved with a third-order WENO method \cite{WENO,ANMPDE}, treating the
electric field, scattering, and quantum terms as source terms; (ii)
the parabolic heat conduction term is incorporated with the TRBDF2
\cite{TRBDF2,ANMPDE} method; and (iii) the elliptic Poisson equation
is solved with a standard elliptic solver---a banded direct solve
(best for 1D) or a modern iterative method \cite{ANMPDE} like PCG or
GMRES.

The time-dependent smooth QHD equations must be regularized to prevent
an unstable growing mode that occurs for barrier heights $B \ge 0.05$
eV for the resonant tunneling diode simulated here.  The simplest
possible regularization involves replacing the spatial derivative of
the electron density on the right-hand side of the momentum
conservation equation by using the classical Boltzmann distribution
for electron density.

The simulations are in good agreement with fully quantum mechanical
simulations (as given by the simulator NEMO \cite{NEMO}) of the
resonant tunneling diode.

This note is an addendum to the author's ``Time-dependent numerical
methods for a regularized quantum hydrodynamic model''
\cite{regular-QHD}.

\section{Smooth QHD Model}
\label{sec-QHDbar}

The classical and quantum hydrodynamic conservation laws (for both the
$O(\hbar^2)$ and smooth QHD models) have the same form\footnote{For
  simplicity, just electrons are treated here. Copies of the transport
  equations (plus transition and generation/recombination terms) can
  be added for holes, upper and lower valley electrons, etc.}:
\begin{equation}
  \frac{\partial}{\partial t} (m n) + 
  \frac{\partial}{\partial x_i} \left(m n u_i\right) = 0,
\label{QHD-n}
\end{equation}
\begin{equation}
  \frac{\partial}{\partial t} \left(m n u_j\right) +
  \frac{\partial}{\partial x_i}	\left(m n u_i u_j - P_{ij}\right) =
  - n \frac{\partial V}{\partial x_j} - \frac{m n u_j}{\tau_p},
\label{QHD-p}
\end{equation}
\begin{equation}
  \frac{\partial W}{\partial t} + \frac{\partial}{\partial x_i}
  \left(u_i W - u_j P_{ij} + q_i\right) =
  - n u_i \frac{\partial V}{\partial x_i} -
  \frac{\left(W_{cl} - \frac{3}{2} n T_0\right)}{\tau_w},
\label{QHD-W}
\end{equation}
where $m$ is the effective electron mass, $n$ is the electron density,
$m n$ is the electron mass density, $u_i$ is the velocity, $m n u_i$
is the momentum density, $P_{ij}$ is the stress tensor, $V$ is the
classical potential energy, $W$ is the total (classical plus quantum)
energy density with the classical energy density given by
\begin{equation}
  W_{cl} = \frac{3}{2} n T + \frac{1}{2} m n u^2,
\end{equation}
$T$ is the temperature of the electron gas, $T_0$ is the ambient
temperature, and $q_i$ is the heat flux.  Boltzmann's constant $k_B$
has been set to 1.  Indices $i$, $j$ equal 1, 2, 3, and repeated
indices are summed over).  Electron-phonon scattering is modeled by
the standard relaxation time approximation, with momentum and energy
relaxation times $\tau_p$ and $\tau_w$.

Equation \eq{QHD-n} expresses conservation of total electron mass,
\eq{QHD-p} expresses conservation of momentum, and \eq{QHD-W}
expresses conservation of energy.

Quantum effects enter into the expressions for the stress tensor
$P_{ij}$ and energy density $W$, which distinguish between the
$O(\hbar^2)$ and smooth QHD models, and in the $O(\hbar^2)$ term in
the generalized heat flux $q_i$ \eq{q}.  The quantum contributions to
the stress tensor and energy density are directly proportional to the
square of the {\em thermal Planck constant}\/
\begin{equation}
  \hbar_\beta^2 \equiv \frac{\hbar^2 \beta}{4 m} = \frac{\hbar^2}{4 m T_0} 
  \approx
  \left\{ \begin{array}{ll}
    \left( 3.42~{\rm nm} \right)^2, & ~T_0 = 300~{\rm K}, \\
    \left( 6.75~{\rm nm} \right)^2, & ~T_0 = 77~{\rm K},
  \end{array} \right.
\end{equation}
where $\beta \equiv 1/T_0$ and the numerical values are for GaAs.  The
quantum contribution to $q_i$ is independent of temperature and
proportional to $\hbar^2$.

The transport equations (\ref{QHD-n})--(\ref{QHD-W}) are coupled to
Poisson's equation for the electrostatic potential energy $V_P$:
\begin{equation}
  \nabla \cdot
  \left(\varepsilon \nabla V_P\right) = e^2 \left(N_D - N_A - n\right), \quad
  V_P = -e \phi, \quad e {\bf E} = \nabla V_P,
\label{QHD-Poisson}
\end{equation}
where $\phi$ is the electrostatic potential, $e > 0$ is the electronic
charge, ${\bf E}$ is the electric field, $\varepsilon$ is the
dielectric constant, $N_D$ is the density of donors, and $N_A$ is the
density of acceptors.

\begin{figure}[htb]
\begin{center}
\scalebox{0.5}{\includegraphics{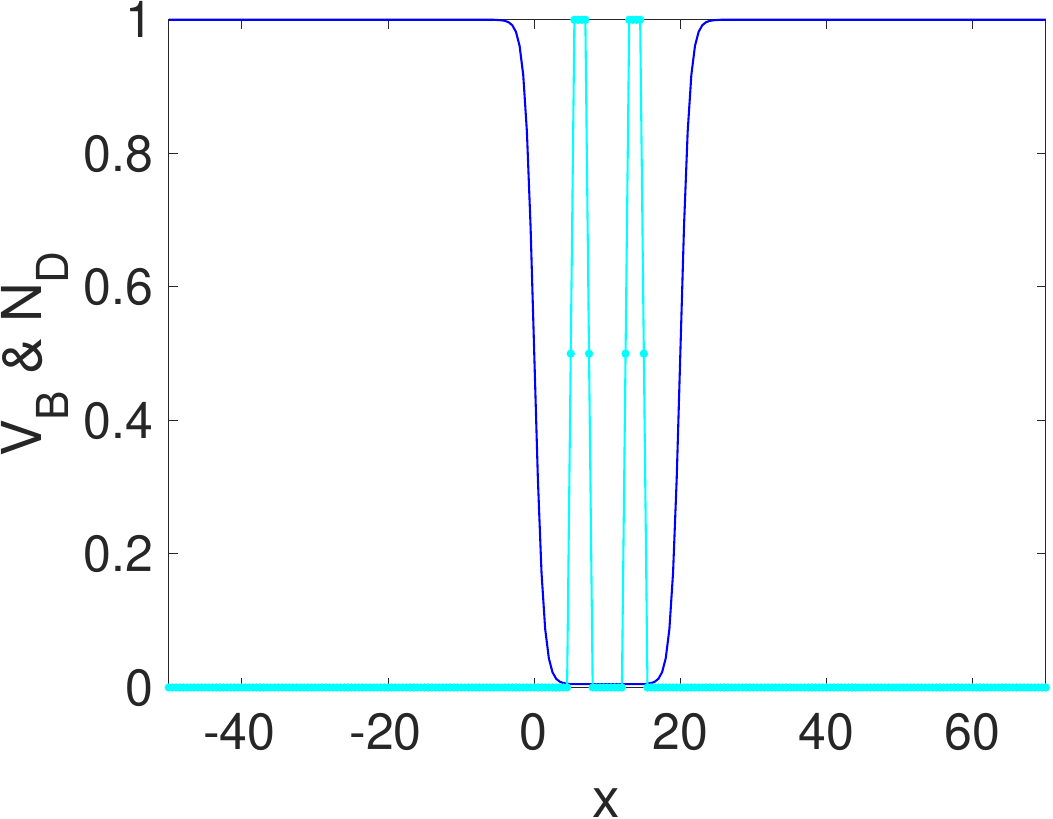}}
\end{center}
\caption{Doping density $N_D$ (blue) in $10^{18}$ cm$^{-3}$ and double
  barriers $V_B/B$ (cyan), where $B$ is the barrier height, for the
  resonant tunneling diode vs.\ $x$ in nm.  The channel is 20 nm long,
  the barriers are 2.5 nm wide, and the quantum well between the
  barriers is 5 nm wide.  There are 5 nm spacers between the barriers
  and contacts.  The dots indicate grid point values for 240
  $\Delta x$.}
\label{fig-NandB}
\end{figure}

The potential energy $V$ consists of two parts, one from the potential
barriers $V_B$ and the other from Poisson's equation $V_P$:
\begin{equation}
  V = V_B + V_P .
\end{equation}
$V_B$ has a step function discontinuity at potential barriers (see
Fig.~\ref{fig-NandB}).  The zero of $V_B$ is uniquely defined by the
{\em absence}\/ of potential barriers.

The relaxation times $\tau_p$ and $\tau_w$ in the QHD transport
equations are given by modified Baccarani-Wordeman \cite{BW} models
\begin{equation}
  \tau_p = m \mu_{n0} \frac{T_0}{T} \equiv \tau_{p0} \frac{T_0}{T}, \quad
  \tau_w = \frac{\tau_p}{2} 
  \left( 1 + \frac{\frac{3}{2} T}{\frac{1}{2} m v_s^2} \right),
\end{equation}
where $\mu_{n0}$ is the low-field electron mobility and $v_s$ is the
electron saturation velocity.  {\em The scattering terms
  $\sim 1/\tau_p$ and $1/\tau_w$ are turned off in the quantum
  region(s) of the device, and are only operative in the semiclassical
  source and drain regions.}

The potential energy types used in this note are listed in Table
\ref{table:potentials}.  The subscript $P$ indicates the electrostatic
Poisson contribution to the potential energy and the subscript $B$
indicates the barrier contribution (see Fig.~\ref{fig-NandB}) to the
potential energy.  In each case, the total potential energy consists
of a Poisson part and a barrier part:
\begin{equation}
  V = V_P + V_B, \quad
  \overline{V} = \overline{V}_P + \overline{V}_B, \quad
  U = U_P + U_B,
\end{equation}
where the {\em quantum potential (energy)}\/ $\overline{V}$ is defined
below in \eq{Vbar} and the {\em smooth potential (energy)}\/ $U$ in
\eq{U}.

\begin{table}[htbp]
  \caption{\label{table:potentials} Potential energy types.}
\begin{center}
\begin{tabular}{l c c} \hline
{\em Type} & {\em \quad Symbol \quad} & {\em \quad Examples \quad} \\ \hline
classical & $V$ & $V_P$, $V_B$ \\
quantum & $\overline{V}$ & $\overline{V}_P$, $\overline{V}_B$ \\
smooth & $U$ & $U_P$, $U_B$ \\ \hline  
\end{tabular}
\end{center}
\end{table}

The stress tensor and energy density for the smooth QHD model are
\cite{SmoothQHD}
\begin{equation}
  P_{ij} = - n T \delta_{ij} - \hbar_\beta^2 n
  \frac{\partial^2 \overline{V}}{\partial x_i \partial x_j},
\label{Pij}
\end{equation}
\begin{equation}
  W = \frac{3}{2} n T + \frac{1}{2} m n u^2 +
  \frac{1}{2} \hbar_\beta^2 n \nabla^2 \overline{V},
\label{W}
\end{equation}
where $n T$ is the pressure $P$ and $\overline{V}$ is the quantum
potential, given by
\begin{displaymath}
  \overline{V}(\beta,{\bf x}) = 
  \int_0^\beta \frac{d\beta'}{\beta} 
  \left( \frac{\beta'}{\beta} \right)^2 
  \int d^3x' \left( \frac{2 m \beta}{\pi (\beta - \beta') (\beta + \beta')
  \hbar^2} \right)^{3/2} ~\times
\end{displaymath}
\begin{equation}
  \exp\left\{ -\frac{2 m \beta}{ (\beta - \beta') (\beta + \beta') \hbar^2} 
  ({\bf x}' - {\bf x})^2 \right\} V({\bf x}') .
\label{Vbar}
\end{equation}
Note the double smoothing of the potential $V$: $\int d\beta'$
integrating over inverse temperature and $\int d^3x'$ integrating over
physical space.

The generalized heat flux \cite{SmoothQHD}
\begin{equation}
  {\bf q} = -\bar{\kappa} n \nabla T - \frac{\hbar^2 n}{8 m} \nabla^2 {\bf u}
\label{q}
\end{equation}
includes a classical Fourier-law term plus a quantum term proportional
to $\hbar^2$, and incorporates significant effects of the higher
moments of the Wigner-Boltzmann equation omitted in the derivation of
the QHD models.  The classical heat flux coefficient $\kappa$ is given
by
\begin{equation}
	\kappa = \kappa_0 \mu_{n0} n T_0 \equiv \bar{\kappa} n,
\end{equation}
where $\mu_{n0}$ is the low-field electron mobility and $\kappa_0$ is
a dimensionless phenomenological constant satisfying
$0 < \kappa_0 \le 2.5$.  On the right-hand side of the QHD model
conservation of energy equation, the generalized heat conduction term
$-\nabla \cdot {\bf q}$ consists of a classical diffusive term
\begin{equation}
  \nabla \cdot (\kappa \nabla T)
\end{equation}
plus a dispersive quantum term
\begin{equation}
  \frac{\hbar^2}{8 m} \nabla \cdot \left(n \nabla^2 {\bf u}\right) .
\end{equation}
The dispersive quantum heat term is handled as a source term in the
third-order WENO method (which uses a third-order Runge-Kutta
time-stepping method).

Plugging in the expressions for the stress tensor \eq{Pij}, energy
density \eq{W}, and heat flux \eq{q}, and setting the smooth potential
\begin{equation}
  U = V + \hbar_\beta^2 \nabla^2 \overline{V},
\label{U}
\end{equation}
the smooth QHD transport equations take the form:
\begin{equation}
  \frac{\partial}{\partial t} (m n) + 
  \frac{\partial}{\partial x_i} \left(m n u_i\right) = 0,
\label{QHD-n2}
\end{equation}
\begin{equation}
  \frac{\partial}{\partial t} \left(m n u_j\right) +
  \frac{\partial}{\partial x_i}	\left(m n u_i u_j\right) +
  \frac{\partial}{\partial x_j} (n T) =
  - n \frac{\partial U}{\partial x_j} -
  \hbar_\beta^2 \frac{\partial n}{\partial x_i}
  \frac{\partial^2 \overline{V}}{\partial x_i \partial x_j} -
  \frac{m n u_j}{\tau_p},
\label{QHD-p2}
\end{equation}
\begin{displaymath}
  \frac{\partial W}{\partial t} + \frac{\partial}{\partial x_i}
  \left(u_i (W + n T) \right) = 
  - n u_i \frac{\partial U}{\partial x_i} -
  \hbar_\beta^2 \frac{\partial (n u_i)}{\partial x_j}
  \frac{\partial^2 \overline{V}}{\partial x_i \partial x_j} ~+ 
\end{displaymath}
\begin{equation}
  \bar{\kappa} \frac{\partial}{\partial x_i}
  \left(n \frac{\partial T}{\partial x_i}\right) +
  \frac{\hbar^2}{8 m} \frac{\partial}{\partial x_i}
  \left(n \nabla^2 u_i\right) -
  \frac{\left(W_{cl} - \frac{3}{2} n T_0\right)}{\tau_w},
\label{QHD-W2}
\end{equation}
with conserved variables mass density $m n$, momentum density
$m n u_j$, and total (classical plus quantum) energy density $W$.

Note that there are no derivatives of the discontinuous potential
$V_B$ (which appears in $V$) in the smooth QHD equations, only of the
smoother potentials $\overline{V}_B$ (in $\overline{V}$) and $U_B$ (in
$U$).

\section{Regularization of the Smooth QHD Model}
\label{sec-regular}

The time-dependent smooth QHD equations must be regularized to prevent
an unstable growing mode that occurs for barrier heights $B \ge 0.05$
eV for the resonant tunneling diode simulated here.

The {\em classical Boltzmann regularization}\/ involves simply
replacing the spatial derivative of the electron density on the
right-hand side of the momentum conservation equation by using the
classical Boltzmann distribution for electron density.

In \cite{regular-QHD} we applied a different {\em quantum Boltzmann
  regularization}---that captured in a more complex way essential
effects of quantum tunneling and resonance---of the time-dependent
smooth QHD model \cite{SmoothQHD} to simulating negative differential
resistance in the resonant tunneling diode.  These were the first
time-dependent smooth QHD simulation results in the literature.  The
splitting method employed in \cite{hyster} for time-dependent
simulations of the original $O(\hbar^2)$ QHD model was modified and
extended to the smooth QHD model, using the third-order WENO method
for the gas dynamical propagation and the implicit TRBDF2 method for
the diffusive heat conduction term, rather than an explicit method,
allowing for larger timesteps.

For time-dependent simulations of the smooth QHD model, the term
\begin{equation}
  - \hbar_\beta^2 \frac{\partial n}{\partial x_i}
  \frac{\partial^2 \overline{V}}{\partial x_i \partial x_j}
\label{regular}
\end{equation}
in the momentum conservation equation \eq{QHD-p2} must be regularized
to prevent an unstable growing mode: the soundspeed squared is given
by \cite{IMA-Disp}
\begin{equation}
  c^2 = \frac{T + \hbar_\beta^2 \nabla^2 \overline{V}}{m} \approx
  \frac{T + \hbar_\beta^2 \nabla^2 \overline{V}_B}{m},
\end{equation}
which becomes negative from the $\nabla^2 \overline{V}_B$ term at some
grid points for barrier heights $B \ge 0.05$ eV ($B/T_0 \ge 2$) for
the resonant tunneling diode simulated here, producing the unstable
growing mode.  The unregularized smooth QHD equations are well-posed
however for small barrier heights ($B/T_0 < 2$), somewhat higher than
the limit ($|V|/T_0 \ll 1$) in which the model was derived.

The growing mode could be controlled by adding viscosity terms to
\eq{QHD-p2} and \eq{QHD-W2}, but then the viscosity makes the NDR too
shallow.  Since viscosity is vanishingly small for quantum
semiconductor devices, we will instead regularize the time-dependent
smooth QHD equations by setting the electron density in \eq{regular}
in the momentum conservation equation \eq{QHD-p2} and in
\begin{equation}
  - \hbar_\beta^2 \frac{\partial (n u_i)}{\partial x_j}
  \frac{\partial^2 \overline{V}}{\partial x_i \partial x_j}
\end{equation}
in the energy conservation equation \eq{QHD-W2} to the classical
Boltzmann distribution ({\em the simplest possible regularization}):
\begin{equation}
  n \sim e^{-V_P/T}, \quad
  n u_i \sim u_i e^{-V_P/T},
\label{n-Boltz0}
\end{equation}
assuming here that $u_i$ is slowly varying\footnote{The form
  \eq{n-Boltz0} for $n u_i$ also maintains the symmetry between the
  right-hand sides of the momentum \eq{QHD-p2} and energy \eq{QHD-W2}
  conservation equations.}.

The dependence of $T$ on ${\bf x}$ is needed in \eq{n-Boltz0} in order
to achieve NDR---this assumption seems quite natural in that the
electron density in the Boltzmann distribution should depend on the
local temperature $T$; if $T_0$ is used, NDR does not occur.  We will
assume though that $T$ is varying slowly enough so that factors of
$\nabla T$ can be ignored below in \eq{nx-Boltz0}.  Keeping terms like
$V_P \nabla T$ would spoil the gauge invariance under
$V_P \rightarrow V_P + const$ of the smooth QHD equations; also
typically $|U \nabla T/T| \ll |\nabla U| $.

On the other hand, we view the term
$\hbar_\beta^2 \partial_i \partial_j \overline{V}$ as an intrinsic
property of the semiconductor barrier/well structure at the ambient
temperature $T_0$, independent of the electron temperature
$T({\bf x},t)$.

Using \eq{n-Boltz0},
\begin{equation}
  \frac{\partial n}{\partial x_i} \approx -\frac{n}{T}
  \frac{\partial V_P}{\partial x_i}, \quad
  \frac{\partial (n u_i)}{\partial x_j} \approx -\frac{n u_i}{T}
  \frac{\partial V_P}{\partial x_j}
\label{nx-Boltz0}
\end{equation}
and the regularized smooth QHD equations become
\begin{equation}
  \frac{\partial}{\partial t} (m n) + 
  \frac{\partial}{\partial x_i} \left(m n u_i\right) = 0,
\label{QHD-n3}
\end{equation}
\begin{equation}
  \frac{\partial}{\partial t} \left(m n u_j\right) +
  \frac{\partial}{\partial x_i}	\left(m n u_i u_j\right) +
  \frac{\partial}{\partial x_j} (n T) =
  -n \frac{\partial U}{\partial x_j} +
  \hbar_\beta^2 \frac{n}{T} \frac{\partial V_P}{\partial x_i}
  \frac{\partial^2 \overline{V}}{\partial x_i \partial x_j} -
  \frac{m n u_j}{\tau_p},
\label{QHD-p3}
\end{equation}
\begin{displaymath}
  \frac{\partial W}{\partial t} + \frac{\partial}{\partial x_i}
  \left(u_i (W + n T) \right) =
  -n u_i \frac{\partial U}{\partial x_i} +
  \hbar_\beta^2 \frac{n u_i}{T} \frac{\partial V_P}{\partial x_j}
  \frac{\partial^2 \overline{V}}{\partial x_i \partial x_j} ~+
\end{displaymath}
\begin{equation}
  \bar{\kappa} \frac{\partial}{\partial x_i}
  \left(n \frac{\partial T}{\partial x_i}\right) +
  \frac{\hbar^2}{8 m} \frac{\partial}{\partial x_i} \left(n
  \nabla^2 u_i\right) -
  \frac{\left(W_{cl} - \frac{3}{2} n T_0\right)}{\tau_w},
\label{QHD-W3}
\end{equation}
\begin{equation}
  \nabla \cdot
  \left(\varepsilon \nabla V_P\right) = e^2 \left(N_D - N_A - n\right), \quad
  V_P = -e \phi, \quad e {\bf E} = \nabla V_P .
\label{Poisson3}
\end{equation}
The right-hand side terms linear in the gradient of the classical
potential energy in the original momentum \eq{QHD-p2} and energy
\eq{QHD-W2} conservation equations have been augmented in \eq{QHD-p3}
and \eq{QHD-W3} by an additional {\em nonlinear}\/ term in the
derivatives of the potential energies:
\begin{equation}
  \frac{\partial V}{\partial x_j} \longrightarrow
  \frac{\partial U}{\partial x_j} -
  \frac{\hbar_\beta^2}{T} \frac{\partial V_P}{\partial x_i}
  \frac{\partial^2 \overline{V}}{\partial x_i \partial x_j} .
\end{equation}

\section{Smooth QHD Simulations of the Resonant Tunneling Diode}
\label{sec-sim}

We will solve the 1D time-dependent smooth QHD equations (with the
classical Boltzmann regularization) to steady-state for the resonant
tunneling diode with the conserved variables mass density $m n$,
momentum density $m n u$, and total energy density $W$:
\begin{equation}
  V = V_B + V_P, \quad
  U \approx U_B + V_P, \quad
  U_B = V_B + \hbar^2_\beta \overline{V}_{Bxx}, \quad
  \overline{V}_{xx} \approx \overline{V}_{Bxx},
\end{equation}
\begin{equation}
  W = \frac{3}{2} n T + \frac{1}{2} m n u^2 +
  \frac{1}{2} \hbar_\beta^2 n \overline{V}_{xx}, \quad
  W_{cl} = \frac{3}{2} n T + \frac{1}{2} m n u^2,
\end{equation}
\begin{equation}
  (m n)_t + (m n u)_x = 0,
\label{QHD-n1}
\end{equation}
\begin{equation}
  (m n u)_t + \left(m n u^2 + n T\right)_x = -n U_x +
  \hbar_\beta^2 \frac{n}{T} V_{Px} \overline{V}_{xx} -
  \frac{m n u}{\tau_p},
\label{QHD-p1}
\end{equation}
\begin{displaymath}
  W_t + (u (W + n T))_x = 
\end{displaymath}
\begin{equation}
  -n u U_x + \hbar_\beta^2 \frac{n u}{T} V_{Px} \overline{V}_{xx} + 
  \bar{\kappa} \left(n T_x\right)_x  +
  \frac{\hbar^2}{8 m} \left(n u_{xx}\right)_x -
  \frac{W_{cl} - \frac{3}{2} n T_0}{\tau_w},
\label{QHD-W1}
\end{equation}
\begin{equation}
  V_{Pxx} = \frac{e^2}{\varepsilon} \left(N_D - n\right), \quad
  V_P = -e \phi, \quad e E =  V_{Px} .
\label{Poisson1}
\end{equation}

Equation (\ref{QHD-n1}) is hyperbolic with characteristic velocity
$u$, equations (\ref{QHD-p1}) and (\ref{QHD-W1}) form a parabolic pair
of equations due to the combination of heat conduction and quantum
terms \cite{IMA-Disp}, and Poisson's equation (\ref{Poisson1}) is
elliptic, so seven boundary conditions are required, four at the left
inflow boundary and three at the right outflow boundary.  Seven
physically relevant, well-posed boundary conditions are
\begin{displaymath}
  n(x_L) = N_D(x_L), \quad u_x(x_{L,R}) = 0, \quad T_x(x_{L,R}) = 0,
\end{displaymath}
\begin{equation}
  V_P(x_L) = 0, \quad V_P(x_R) = -e V_{bias},
\end{equation}
where $x_L$ and $x_R$ denote the left and right boundaries,
respectively, of the resonant tunneling diode.

\subsection{Simulations of Negative Differential Resistance}
\label{sec-NDR}

\begin{figure}[htb]
\begin{center}
\scalebox{0.5}{\includegraphics{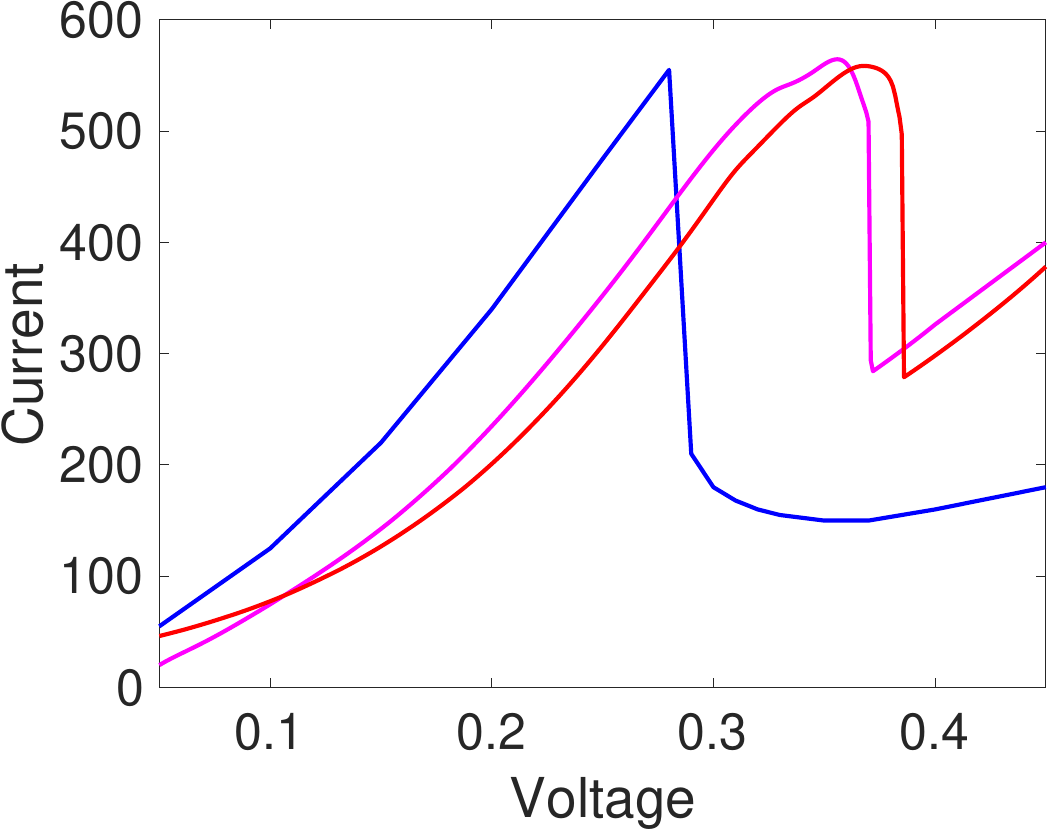}}
\caption{Current (density) in kA/cm$^2$ vs.\ applied voltage in volts
  for the resonant tunneling diode at 300~K with a 5 nm quantum well
  and 2.5 nm 0.28 eV quantum barriers for the classical (red) and
  quantum (magenta) Boltzmann regularizations showing realistic NDR,
  compared with the NEMO simulation (blue) by G.~Klimeck \cite{NEMO}.}
\label{fig-IVs}
\end{center}
\end{figure}

The smooth QHD model using the quantum Boltzmann regularization was
shown in \cite{regular-QHD} to give good qualitative agreement with
the current-voltage curves for the resonant tunneling diode predicted
by the simulator NEMO \cite{NEMO}, which is based on solving the full
mixed state quantum mechanics via a nonequilibrium Green's function
technique.  We will demonstrate here that simulations using the
classical Boltzmann regularization \eq{n-Boltz0} and \eq{nx-Boltz0} of
the smooth QHD model still match fully quantum mechanical simulations
(cf.\ the NEMO results in Fig.~\ref{fig-IVs} for the exact same
device) of the resonant tunneling diode.

\begin{figure}[htb]
\begin{center}
\scalebox{0.5}{\includegraphics{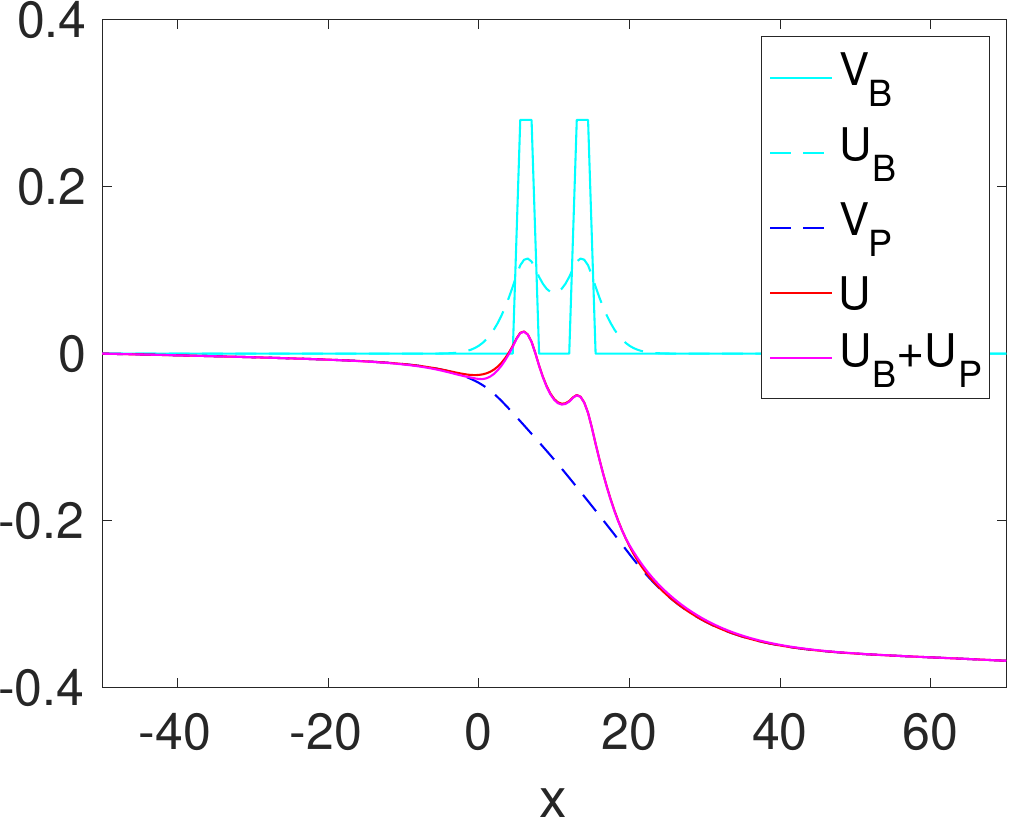}}
\end{center}
\caption{The smooth potential $U \approx U_B + V_P$ in eV for the
  resonant tunneling diode at the resonant peak current with
  $V_{bias} = 370$ mV and $B$ = 0.28 eV vs.\ $x$ in nm, compared with
  the exact smooth potential $U_B + U_P$.}
\label{fig-Upeak}
\end{figure}

\begin{figure}[htb]
\begin{center}
\scalebox{0.5}{\includegraphics{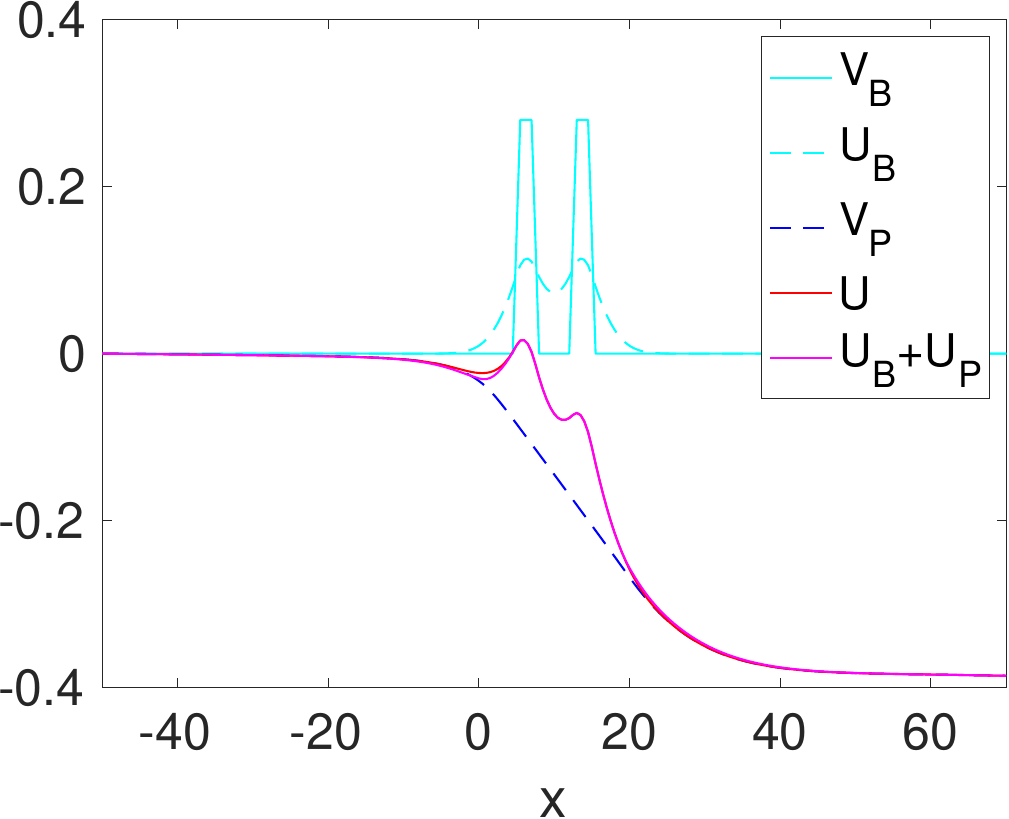}}
\end{center}
\caption{The smooth potential $U \approx U_B + V_P$ in eV for the
  resonant tunneling diode at the current valley with $V_{bias} = 385$
  mV and $B$ = 0.28 eV vs.\ $x$ in nm, compared with the exact smooth
  potential $U_B + U_P$.}
\label{fig-Uvalley}
\end{figure}

\begin{figure}[htb]
\begin{center}
\scalebox{0.5}{\includegraphics{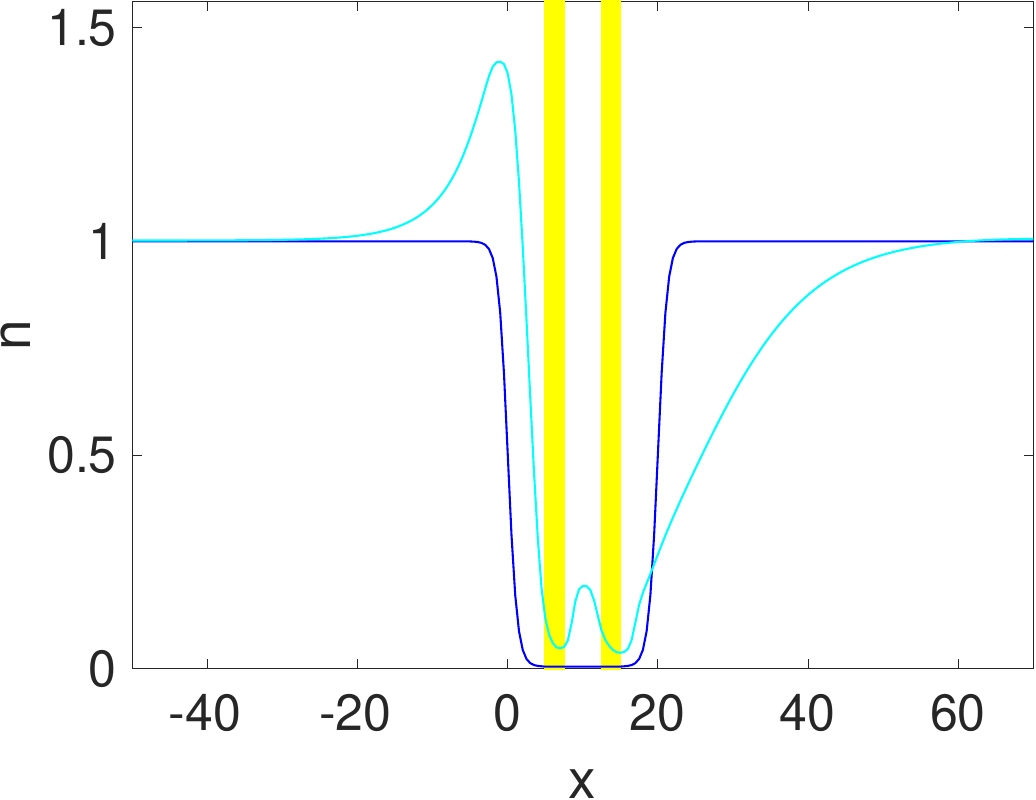}}
\end{center}
\caption{Electron density $n$ (cyan) and, for reference, doping
  density $N_D$ (blue) in $10^{18}$ cm$^{-3}$ vs.\ $x$ in nm in the
  resonant tunneling diode at the resonant peak current with
  $V_{bias} = 370$ mV and $B$ = 0.28 eV\@.}
\label{fig-npeak}
\end{figure}

\begin{figure}[htb]
\begin{center}
\scalebox{0.5}{\includegraphics{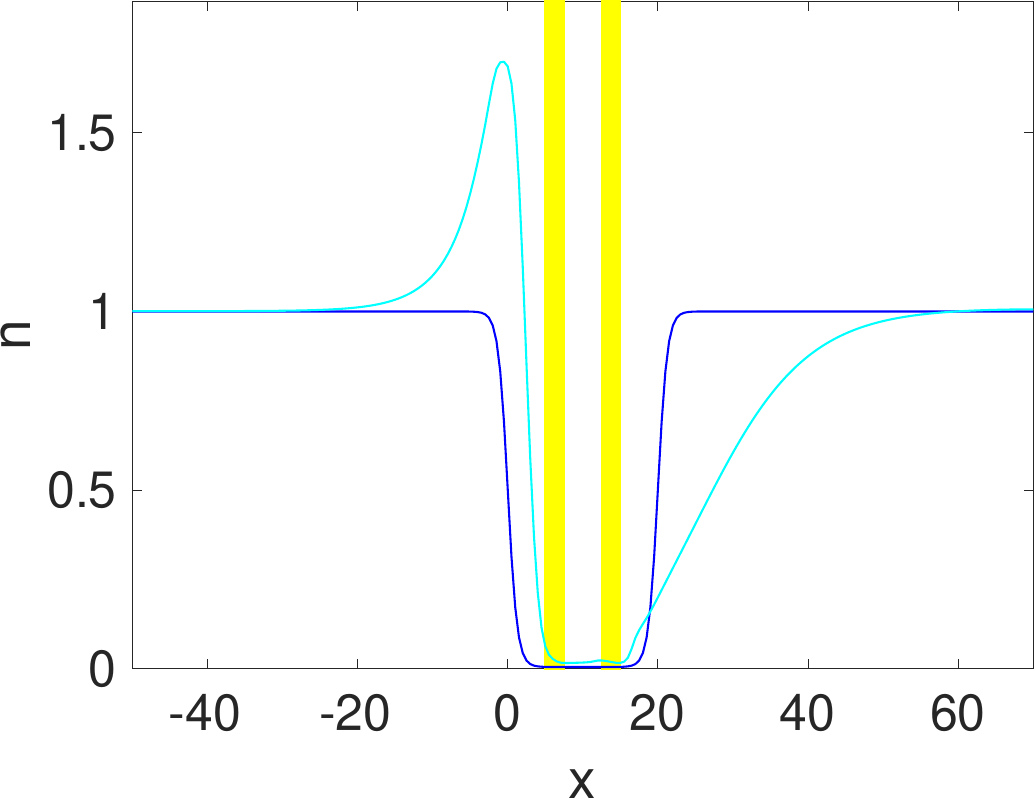}}
\end{center}
\caption{Electron density $n$ (cyan) and, for reference, doping
  density $N_D$ (blue) in $10^{18}$ cm$^{-3}$ vs.\ $x$ in nm in the
  resonant tunneling diode at the valley current with $V_{bias} = 385$
  mV and $B$ = 0.28 eV\@.}
\label{fig-nvalley}
\end{figure}

\begin{figure}[htb]
\begin{center}
\scalebox{0.5}{\includegraphics{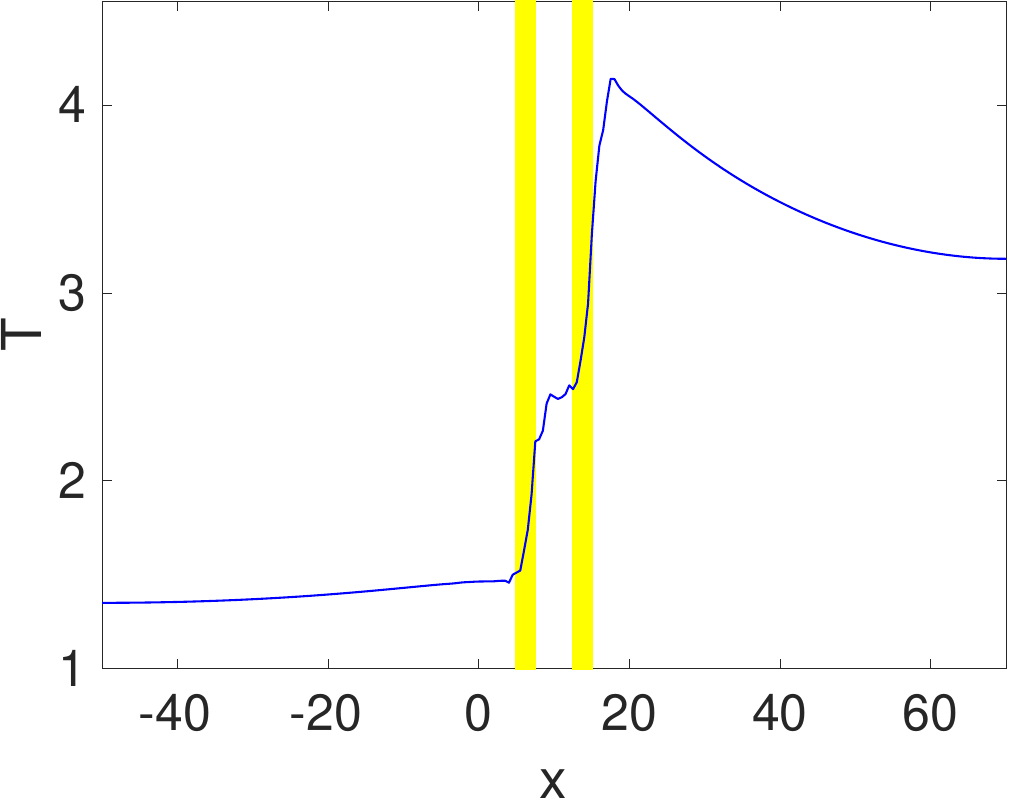}}
\end{center}
\caption{Temperature $T/T_0$ vs.\ $x$ in nm in the resonant tunneling
  diode at the resonant peak current with $V_{bias} = 370$ mV and $B$ =
  0.28 eV\@.}
\label{fig-Tpeak}
\end{figure}

Smooth QHD model simulations with 240 $\Delta x$ of a GaAs resonant
tunneling diode at 300~K with $B$ = 0.28 eV AlGaAs double barriers are
presented in Figs.~\ref{fig-IVs}--\ref{fig-U-in-de}.  The diode
consists of an $n^+$ source (at the left) and drain (at the right),
each with doping $N_D = 10^{18}$ cm$^{-3}$, and an $n$ channel with
doping $N_D = 5 \times 10^{15}$ cm$^{-3}$ (see Fig.~\ref{fig-NandB}).
The channel is 20 nm long, the barriers are 2.5 nm wide, and the
quantum well between the barriers is 5 nm wide.  There are 5 nm
spacers between the barriers and contacts.

Physical parameters for GaAs at 300~K are electron effective mass
$m = 0.063 \, m_e$, saturation velocity $v_s = 1.5 \times 10^7$ cm/s,
and dielectric constant $\varepsilon = 12.9$.  The heat flux
pre-factor is set to the canonical value $\kappa_0 = 1.5$, while the
relaxation time is adjusted slightly from the NEMO calculated value
\cite{NEMO} of $\tau_{p0} = 0.07$ ps to $\tau_{p0} = 0.09$ ps to
produce the most realistic NDR in Fig.~\ref{fig-IVs}.

For the current-voltage curve of the resonant tunneling diode, we
simulate the time-dependent equations to steady state for each applied
voltage $V_{bias}$ across the device, using continuation in
$V_{bias}$.

The figures display the simulations of the resonant tunneling diode
with the classical Boltzmann regularization $n \sim e^{-V_P/T}$
\eq{n-Boltz0} in the $\hbar^2_\beta n_x \overline{V}_{xx}$ and
$\hbar^2_\beta (n u)_x \overline{V}_{xx}$ terms on the right-hand
sides of the smooth QHD momentum and energy conservation equations.

The experimental signal of quantum resonance in the resonant tunneling
diode is negative differential resistance in the current-voltage
curve.

% NEMO peak/valley at V = 0.280-0.270(DD)/0.340, j = 550-450(DD)/150

For the classical Boltzmann regularization, note that the smooth QHD
model simulation correctly predicts significant NDR in the resonant
tunneling diode at 300~K, unlike the original $O(\hbar^2)$ QHD model.
The smooth QHD peak (560 kA/cm$^2$ at 370 mV) and valley (280
kA/cm$^2$ at 385 mV) currents are in good agreement with the NEMO
\cite{NEMO} values (550 kA/cm$^2$ peak current at 280 mV and 200
kA/cm$^2$ valley current at 290 mV), as well as the overall shape of
the current-voltage curve.  The valley of the NEMO current-voltage
curve however is deeper and broader---this could be a modeling
issue\footnote{In $\tau_p$, $\tau_w$, and/or $\kappa$.} or may be a
downside of the fluid dynamical approach.  An exact comparison between
NEMO and the smooth QHD model is impossible though, since NEMO assumes
the contacts are thermal equilibrium reservoirs and the quantum
regions of the device are non-equilibrium, while the smooth QHD
model---as a fluid dynamical approximation---assumes the whole device
is everywhere locally near thermal equilibrium.

The resonant peak of the current-voltage curve at 370 mV occurs as the
electrons tunneling through the first barrier come into resonance with
the ground state energy level of the quantum well, creating enhanced
charge in the quantum well (Figs.~\ref{fig-Upeak} and
\ref{fig-npeak}).  As the voltage bias increases above 370 mV, the
resonance effect rapidly decreases and then disappears because the
effective right barrier height in $U$ in Fig.~\ref{fig-Upeak} is
progressively lowered (Fig.~\ref{fig-Uvalley}), depleting the charge
in the quantum well (Fig.~\ref{fig-nvalley}).

Figure~\ref{fig-Tpeak} illustrates that the electron gas heats up
inside the double barrier structure to approximately $4 T_0$ and then
decays back down to approximately $3.25 T_0$ in the drain.  There is
some residual visible effect of the barriers on the temperature in the
double barrier structure, due to the jump condition \cite{SmoothQHD}:
\begin{equation}
    \bar{\kappa} \left[T_x\right] = -\frac{u}{2} \left[V_B\right],
\end{equation}
where $[\chi] = \chi_+ - \chi_-$ is the jump in a quantity $\chi$
across a barrier edge.  In other words, $T$ is continuous but $T_x$ is
discontinuous at the barrier edges.

\clearpage

\subsection{Simulation of Hysteresis}

The original QHD model succeeded in simulating hysteresis
\cite{hyster} in the current-voltage curve of the resonant tunneling
diode, although the hysteresis loop was smaller than those
experimentally measured.  NDR is suppressed in the original QHD model
at 300~K \cite{IMA-Disp}.

In \cite{hyster-300}, we demonstrated that the smooth QHD model
successfully produces hysteresis in the resonant tunneling diode at
300~K, with a more realistic hysteresis loop.  This was the first
quantum hydrodynamic simulation of hysteresis at 300~K\@.  The smooth
QHD model used there, however, predicted a peak current at 140 eV, so
we recommend instead the new classical or quantum Boltzmann
regularized models of Section \ref{sec-NDR}.

Here we will simulate hysteresis in the resonant tunneling diode at
300~K using the smooth QHD model with the classical Boltzmann
regularization.  Simulating the 1D time-dependent equations to
steady-state, hysteresis at a voltage bias $V_{bias}$ occurs when
different initial states at $V_{bias} - \Delta V$ (voltage increasing)
and $V_{bias} + \Delta V$ (voltage decreasing) produce different
steady-state currents at $V_{bias}$.

\begin{figure}[htb]
\begin{center}
\scalebox{0.5}{\includegraphics{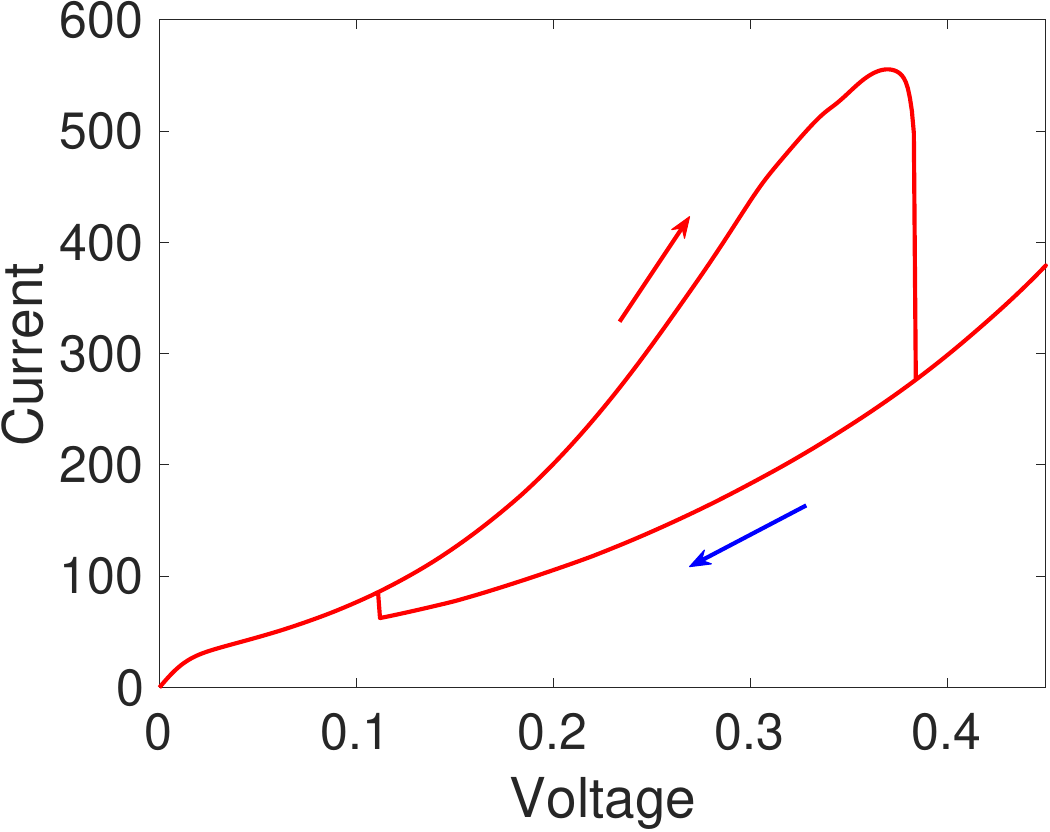}}
\caption{Current in kA/cm$^2$ vs.\ applied voltage in volts showing
  hysteresis for the resonant tunneling diode at 300~K with a 5 nm
  quantum well and 2.5 nm 0.28 eV quantum barriers.  The red arrow
  indicates current for increasing voltage and the blue arrow for
  decreasing voltage.}
\label{fig-IV-hyster}
\end{center}
\end{figure}

Figure~\ref{fig-IV-hyster} displays the experimental signals of
quantum resonance: negative differential resistance and hysteresis in
the current-voltage curve.  Note that the smooth QHD simulation
correctly predicts significant NDR and hysteresis in the resonant
tunneling diode at 300~K, with a more realistic hysteresis loop than
the original QHD \cite{hyster} or quantum drift-diffusion models
\cite{Jungel1,AJ} predicted at 77~K\@.

\begin{figure}[htb]
\begin{center}
\scalebox{0.5}{\includegraphics{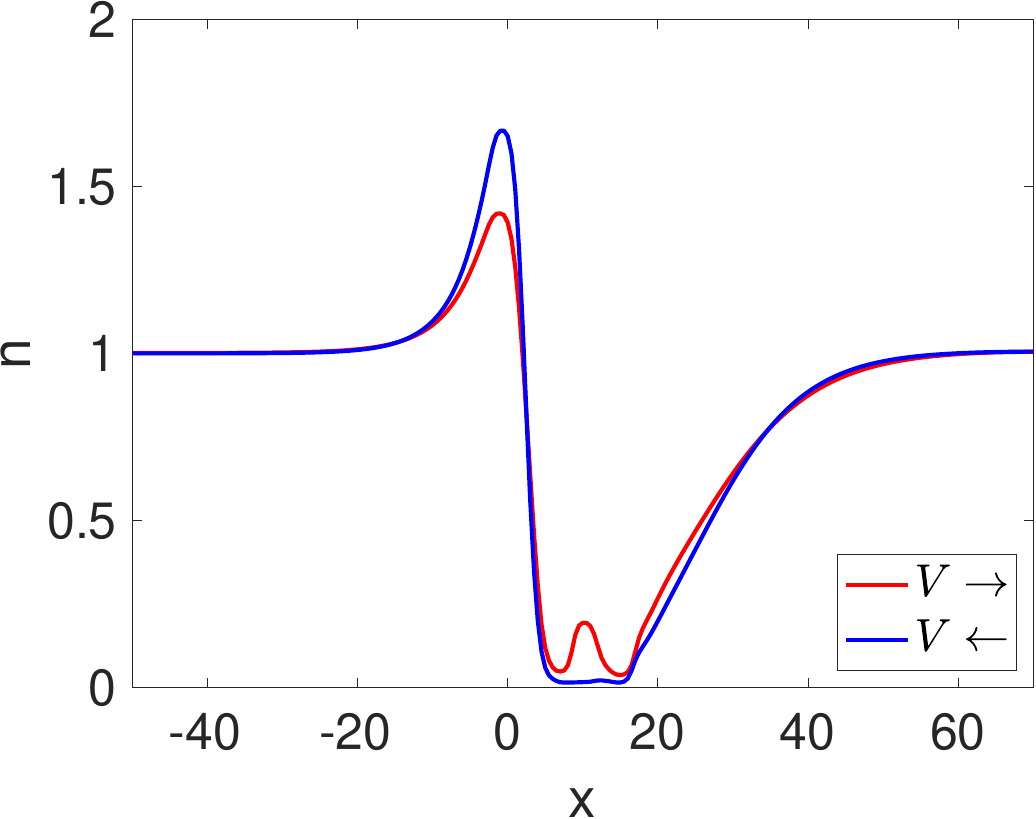}}
\caption{Electron density $n$ in $10^{18}$ cm$^{-3}$ vs.\ $x$ in nm in
  the resonant tunneling diode at the resonant peak voltage $V_{bias}
  = 370$ mV (red increasing/blue decreasing voltage).}
\label{fig-n-in-de}
\end{center}
\end{figure}

\begin{figure}[htb]
\begin{center}
\scalebox{0.5}{\includegraphics{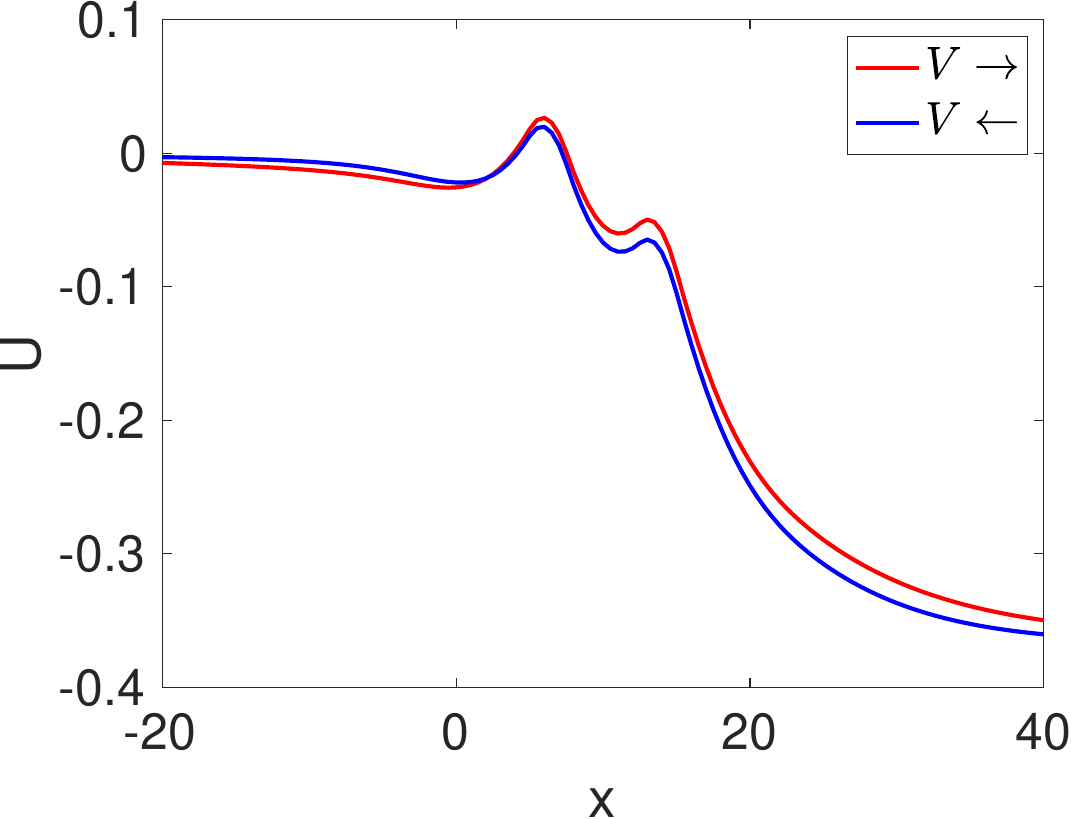}}
\caption{Close-up of the smooth potential $U = V_P + U_B$ in eV for
  the resonant tunneling diode at the resonant peak voltage $V_{bias}
  = 370$ mV (red increasing/blue decreasing voltage) vs.\ $x$ in nm.}
\label{fig-U-in-de}
\end{center}
\end{figure}

The resonant peak of the current-voltage curve at 370 mV occurs as the
electrons tunneling through the first barrier come into resonance with
the energy levels of the quantum well (see Fig.~\ref{fig-n-in-de}
showing enhanced charge in the quantum well).  As the voltage bias
increases above 370 mV, the resonance effect rapidly decreases and
then disappears because the effective right barrier height in $U$ in
Fig.~\ref{fig-U-in-de} is progressively lowered.

The physical mechanism for hysteresis is that electrons ``see'' a
slightly different smooth potential energy $U$ as the applied voltage
is increasing versus decreasing (Fig.~\ref{fig-U-in-de}).  When the
applied voltage is increasing in the hysteresis loop, the effective
quantum well is closer to a parabolic well, producing more resonance
(the increased electron charge in the quantum well in
Fig.~\ref{fig-n-in-de}) and thus amplifying current flow through the
device.

These simulations demonstrate that a fluid dynamical model can
reproduce this fundamental quantum mechanical effect of hysteresis.

In summary, the smooth QHD model (with the classical Boltzmann
regularization) provides an efficient way to incorporate quantum
effects like resonant tunneling, negative differential resistance, and
hysteresis into semiconductor device simulation without having to
solve the much more computationally expensive Wigner-Boltzmann
transport equation or the full mixed-state quantum mechanics.


\clearpage

\begin{thebibliography}{10}

\bibitem{AJ}
{\sc A.~Arnold and A.~J\"ungel}, {\em Multi-scale modeling of quantum
  semiconductor devices}, in Analysis, Modeling and Simulation of Multiscale
  Problems, Springer, 2006, pp.~331--363.

\bibitem{BW}
{\sc G.~Baccarani and M.~R. Wordeman}, {\em An investigation of steady-state
  velocity overshoot effects in {S}i and {G}a{A}s devices}, Solid State
  Electronics, 28 (1985), pp.~407--416.

\bibitem{TRBDF2}
{\sc R.~E. Bank, W.~M. Coughran, W.~Fichtner, E.~H. Grosse, D.~J. Rose, and
  R.~K. Smith}, {\em Transient simulation of silicon devices and circuits},
  {IEEE} Transactions on Computer-Aided Design, 4 (1985), pp.~436--451.

\bibitem{hyster}
{\sc Z.~Chen, B.~Cockburn, C.~L. Gardner, and J.~W. Jerome}, {\em Quantum
  hydrodynamic simulation of hysteresis in the resonant tunneling diode},
  Journal of Computational Physics, 117 (1995), pp.~274--280.

\bibitem{hyster-300}
{\sc C.~L. Gardner}, {\em Quantum hydrodynamic simulation of hysteresis in the
  resonant tunneling diode at 300 {K}}, Journal of Computational Electronics,
  20 (2021), pp.~230--236.

\bibitem{ANMPDE}
{\sc C.~L. Gardner}, {\em Applied Numerical Methods for Partial Differential
  Equations}, vol.~78, Texts in Applied Mathematics, Springer, 2024.

\bibitem{regular-QHD}
{\sc C.~L. Gardner}, {\em Time-dependent numerical methods for a regularized
  quantum hydrodynamic model}, Communications on Applied Mathematics and
  Computation,  (2026).
\newblock //doi.org/10.1007/s42967-026-00598-3.

\bibitem{NEMO}
{\sc C.~L. Gardner, G.~Klimeck, and C.~Ringhofer}, {\em Smooth quantum
  hydrodynamic model vs.\ {NEMO} simulation of resonant tunneling diodes},
  Journal of Computational Electronics, 3 (2004), pp.~95--102.

\bibitem{SmoothQHD}
{\sc C.~L. Gardner and C.~Ringhofer}, {\em Smooth quantum potential for the
  hydrodynamic model}, Physical Review E, 53 (1996), pp.~157--167.

\bibitem{IMA-Disp}
{\sc C.~L. Gardner and C.~Ringhofer}, {\em Dispersive/hyperbolic hydrodynamic
  models for quantum transport (in semiconductor devices)}, in {IMA} {V}olumes
  in {M}athematics and its {A}pplications, vol.~136, Springer, 2003,
  pp.~91--106.

\bibitem{Jungel1}
{\sc A.~J\"ungel and S.~Tang}, {\em Numerical approximation of the viscous
  quantum hydrodynamic model for semiconductors}, Applied Numerical
  Mathematics, 56 (2006), pp.~899--915.

\bibitem{WENO}
{\sc C.-W. Shu}, {\em High order {ENO} and {WENO} schemes for computational
  fluid dynamics}, in High-Order Methods for Computational Physics, vol.~9 of
  Lecture Notes in Computational Science and Engineering, Springer, 1999,
  pp.~439--582.

\end{thebibliography}
\end{document}